\documentclass[twocolumn,prl,aps,superscriptaddress,nofootinbib]{revtex4}

\usepackage{graphicx}
\usepackage{bbold}
\usepackage{amsmath}
\usepackage{float}
\usepackage{color}
\usepackage{bm}

\def\be{\begin{eqnarray}}
\def\ee{\end{eqnarray}}
\def\ben{\begin{eqnarray*}}
\def\een{\end{eqnarray*}}
\def\bes{\begin{subequations}}
\def\ees{\end{subequations}}

\newcommand{\wig}[1]{\mathrel{\hbox{\hbox to 0pt{\lower.6ex\hbox{$\sim$}\hss    }\raise.4ex\hbox{$#1$}}}}

\begin{document}

\title{First-Principles Determination of Electron-Ion Couplings\\ in the Warm Dense Matter Regime}

\author{Jacopo \surname{Simoni}}
\author{J\'er\^ome \surname{Daligault}}
\email{jsimoni@lanl.gov,daligaul@lanl.gov}
\affiliation{Theoretical Division, Los Alamos National Laboratory, Los Alamos, NM 87545, USA}

\begin{abstract}
We present first-principles calculations of the rate of energy exchanges between electrons and ions in nonequilibrium warm dense plasmas, liquid metals and hot solids, a fundamental property for which various models offer diverging predictions.
To this end, a Kubo relation for the electron-ion coupling parameter is introduced, which includes self-consistently the quantum, thermal, non-linear and strong coupling effects that coexist in materials at the confluence of solids and plasmas.
Most importantly, like other Kubo relations widely used for calculating electronic conductivities, the expression can be evaluated using quantum molecular dynamics simulations.
Results are presented and compared to experimental and theoretical predictions for representative materials of various electronic complexity, including aluminum, copper, iron and nickel.
\end{abstract}

%\pacs{...} %52.25.Fi,52.27.Gr,51.20.+d,52.65.−y

\date{\today}

\maketitle

The last decade has seen remarkable progress in our ability to form and interrogate in the laboratory materials under conditions at the confluence of solids and hot plasmas in the so-called warm dense matter regime \cite{Grazianietal2014,Frontierreport2017}.
These experimental advances severely challenge our arsenal of theoretical techniques, simulation tools and analytical models.
In addition to including the coexisting quantum, thermal, disorder and strong Coulomb interaction effects, theoretical approaches are needed that can also describe non-equilibrium conditions \cite{Ng_et_al_1995,Faustlin2010,Ng2012,Chen2013,Leguay2013,Clerouin2015,Zylstra2015,Cho_et_al_2015,Dorchies2016,Jourdain_et_al_2018,Ogitsu2018,Daligault_Dimonte_2009}.
A particularly important property is the electron-ion coupling factor that measures the rate of energy exchanges between electrons and ions \cite{Ng2012}.
Indeed, experiments typically produce transient, non-equilibrium conditions and measurements may be misleading if recorded while the plasma species are still out of equilibrium.
Moreover, like the electron-phonon coupling, the electron-ion coupling may be a unique indicator of the underlying electronic structure and of the basic interaction processes occuring in the warm dense matter regime.
Remarkably, while even for simple materials various models offer diverging predictions (see table~\ref{table_1}), the electron-ion coupling factor is now accessible to experimental measurements thanks to the diagnostic capabilities offered by the new generation of x-ray light sources \cite{Leguay2013,Cho_et_al_2015,Dorchies2016,Jourdain_et_al_2018,Ogitsu2018}.

Here, we use a combination of first-principles theory and ab-initio molecular dynamics simulations to calculate the electron-ion coupling of materials under warm dense matter conditions.
In the same way as with the now routine ab-initio calculations of electrical and thermal conductivities \cite{Desjarlais_et_al_2002, Holst2011,Sjostrom2015}, the approach offers a very useful comparison with the experimental measurements and a useful test of theories, it gives insight into the underlying physics, and it permits an extension into conditions not covered by the experiments. 
The electron-ion coupling is related to the friction coefficients felt by individual ions due to their non-adiabatic interactions with electrons.
Each coefficient satisfies a Kubo relation given by the time integral of the autocorrelation function of the interaction force  of an ion with the electrons, which is evaluated using density functional theory based quantum molecular dynamics simulations.
In this Letter, we outline the underlying theory and present results for a set of relevant materials and physical conditions.
Details of mathematical proofs and algorithms will be presented in an extended manuscript \cite{SimoniDaligault2019}.
Below, $\hbar$ is the reduced Planck constant, $k_B$ is the Boltzmann constant.

\begin{table}[b]
\begin{tabular}{c|c}
Theoretical model & $G_{ei}$ $(10^{17}\,{\rm  W/m^{3}K)}$\\\hline \hline
Spitzer-Brysk & $160$ \cite{Brysk1975}\\
Fermi golden rule & $5$ \cite{Hazak2001,Daligault_Mozyrsky_2008}\\
Coupled modes & $0.33$ \cite{DharmaWardana2001}  ; $0.1$ \cite{Vorberger2012} \\\hline
Electron-phonon & $2.6$ \cite{Lin_et_al_2008}  ; $5$ \cite{Waldecker2016}%\\\hline
\end{tabular}
\caption{Electron-ion coupling for solid density aluminum at melting conditions.
\label{table_1}}
\end{table}
We consider a material of volume $V$ containing one atomic species.
The material is described as a two-component system comprised of ions (mass $m_i=Am_u$, number density $n_i=N_i/V$, charge $Ze$) and of electrons (mass $m_e$, density $n_e=Zn_i$), where each ion consists of an atomic nucleus and its most tightly bound, unresponsive core electrons.
We assume that the material can be described as an isolated, homogeneous, two-temperature system characterized at all times $t$ by the temperatures $T_e(t)$ and $T_i(t)$ of the electronic (e) and ionic (i) subsystems.
Under the mild assumptions recalled below, the temperatures can be shown to evolve according to
\be
c_i^0\frac{dT_i}{dt}=G_{ei}\,(T_e-T_i)\quad,\quad c_e\frac{dT_e}{dt}=-G_{ei}\,(T_e-T_i) \label{ttm_equations}
\ee
where $c_i^0=3n_ik_B/2$ is the ionic kinetic contribution to the heat capacity and $c_{e}$ is the specific heat capacity at constant volume, and
\be
G_{ei}(T_e,T_i)= 3n_i k_B\left\langle \frac{1}{3N_i}\sum_{I=1}^{N_{i}}{\sum_{x=1}^{3}{\gamma_{Ix,Ix}\left({\bf R},T_e\right)}} \right\rangle_{i}\,, \label{Gei}
\ee
is the electron-ion coupling, the focus of this work, given by the thermal average over ionic configurations ${\bf R}=({\bf R}_1,\dots,{\bf R}_{N_i})$ at temperature $T_i$ of the sum over all ions and spatial dimensions of the electron-ion friction coefficient $\gamma_{Ix}\left({\bf R},T_e\right)$ felt by ion $I$ along the $x$-direction as a result of non-adiabatic interactions with the electrons.
The friction coefficients satisfy the Kubo relation
\be
\hspace{-0.2cm}\gamma_{Ix,Jy}\left({\bf R},T_e\right)\!\!&=&\!\!\frac{1}{2m_ik_BT_e}{\rm Re}\int_0^\infty{\!\!dt \left\langle \hat{{\cal{F}}}_{Ix}(t) \hat{{\cal{F}}}_{Jy}(0)\right\rangle_{\!e}} \label{gamma_IxJy_exact}
\ee
where $\left\langle \dots\right\rangle_{\!e}$ is the electronic thermal average at temperature $T_e$, and $\hat{{\cal{F}}}_{Ix}(t)=-e^{i\hat{H}_et/\hbar}\frac{\partial \hat{H}_e\left({\bf R}\right)}{\partial R_{Ix}}e^{-i\hat{H}_et/\hbar}$ is the electron-ion force operator at time $t$,  where $\hat{H}_e\left({\bf R}\right)=\sum_i{\frac{\hat{{\bf p}}_i^2}{2m_e}}+\sum_{i,I}{v_{ie}(\hat{{\bf r}}_i-{\bf R}_{I})}+\sum_{i\neq j}{\frac{e^2}{4\pi\epsilon_0}\frac{1}{|\hat{{\bf r}}_i-\hat{{\bf r}}_j|}}$ is the electronic Hamiltonian.
Here, for simplicity of exposition, the electron-ion interaction is described by a local pseudopotential $v_{ie}(r)$; in practice, the formalism allows to deal with more elaborate descriptions \cite{SimoniDaligault2019} (e.g., the results shown below for noble and transition metals were obtained using plane-augmented wave pseudopotentials).

Equations (\ref{ttm_equations})-(\ref{gamma_IxJy_exact}) result from a first-principles derivation under the following three assumptions \cite{Daligault_Mozyrsky_2009,Daligault_Mozyrsky_2018}.
(i) The dynamics of each ion can be described by that of the center ${\bf R}_i(t)$ of its narrowly localized wavepacket.
This is justified here, since the thermal de Broglie wavelength $\Lambda=\hbar\sqrt{2\pi/m_ik_BT_i}$ ($\simeq 0.3/\sqrt{AT_{i}[eV]}$ Bohr) of ions is generally much smaller than the spatial variations of forces acting on them due to their large mass and the relatively high temperatures.
(ii) The typical ionic velocities are small compared to the typical electronic velocities.
For instance, we assume $T_i/m_i\ll T_F/m_e$ or $T_i/m_i\ll T_e/m_e$ in the degenerate $T_e/T_F\ll 1$ or non-degenerate limit $T_e/T_F\ll 1$, respectively, where $T_F=\frac{\hbar^2}{2m_e k_B}(3\pi^2 n_e)^{\frac{2}{3}}$ ($\simeq 1.69\,\left(n_e[{\rm cm}^{-3}]/10^{22}\right)^{\frac{2}{3}}$ eV) is the electronic Fermi temperature.
This condition is generally respected due to the natural smallness of $m_e/m_i$, and is challenged only if $T_i\gg T_e$.
(iii) Finally, we assume that there is a quasi-continuum of electronic states, as is the case for the metallic systems of interest here.
Under these conditions, the ion dynamics follows the stochastic, Langevin-like equation $m_i\ddot{\bf R}={\bf F}_{BO}+m_i\tensor{\gamma}\cdot\dot{\bf R}+\boldsymbol{\xi}$, and Eqs.(\ref{ttm_equations})-(\ref{gamma_IxJy_exact}) are obtained from the equation of evolution of the ionic energy that results from it \cite{Daligault_Mozyrsky_2009,SimoniDaligault2019}.
Here ${\bf F}_{BO}$ is the adiabatic Born-Oppenheimer force, which includes the interactions between ions and with the instantaneous electrostatic potential of electrons.
The other terms describe the effect of non-adiabatic transitions between closely spaced electronic states induced by the atomic motions and electronic excitations.
These terms, which are not accounted for in current quantum molecular dynamics simulations, are responsible for the constant, non-reversible, energy exchanges between electron and ions.
Like the buffeting of light liquid particles on a heavy Brownian particle, the non-adiabatic effects produce a friction force $M\tensor{\gamma}\cdot\dot{\bf R}$, where $\tensor{\gamma}=\left\{\gamma_{Ix,Jy}\right\}$, and a $\delta$-correlated Gaussian random force $\boldsymbol{\xi}$ with correlator $\prec \xi_{Ix}(t)\xi_{Jy}(t')\succ=2m_ek_BT_e\gamma_{Ix,Jy}\delta(t-t')$.

The expression (\ref{Gei}) includes self-consistently the non-ideal, quantum and thermal effects that coexist in the warm dense matter regime.
It reduces to well-known models in limiting cases \cite{Daligault_Mozyrsky_2008}, including the traditional Spitzer-Brysk formula in the hot plasma limit \cite{Brysk1975} and the Fermi golden rule formula in the limit of weak electron-ion interactions \cite{Hazak2001,Daligault_Mozyrsky_2008}.
Moreover, it applies to hot solids with lattice temperature $T_i$ much larger than the Debye temperature $\Theta_D$ (typically $0.01-0.04$ eV \cite{AshcroftMerminbook}), where it extends the standard electron-phonon coupling $G_{e,ph}$ \cite{Allen1987} by including ionic motions beyond the harmonic approximation.

\begin{figure}[t]
\includegraphics[scale=0.36]{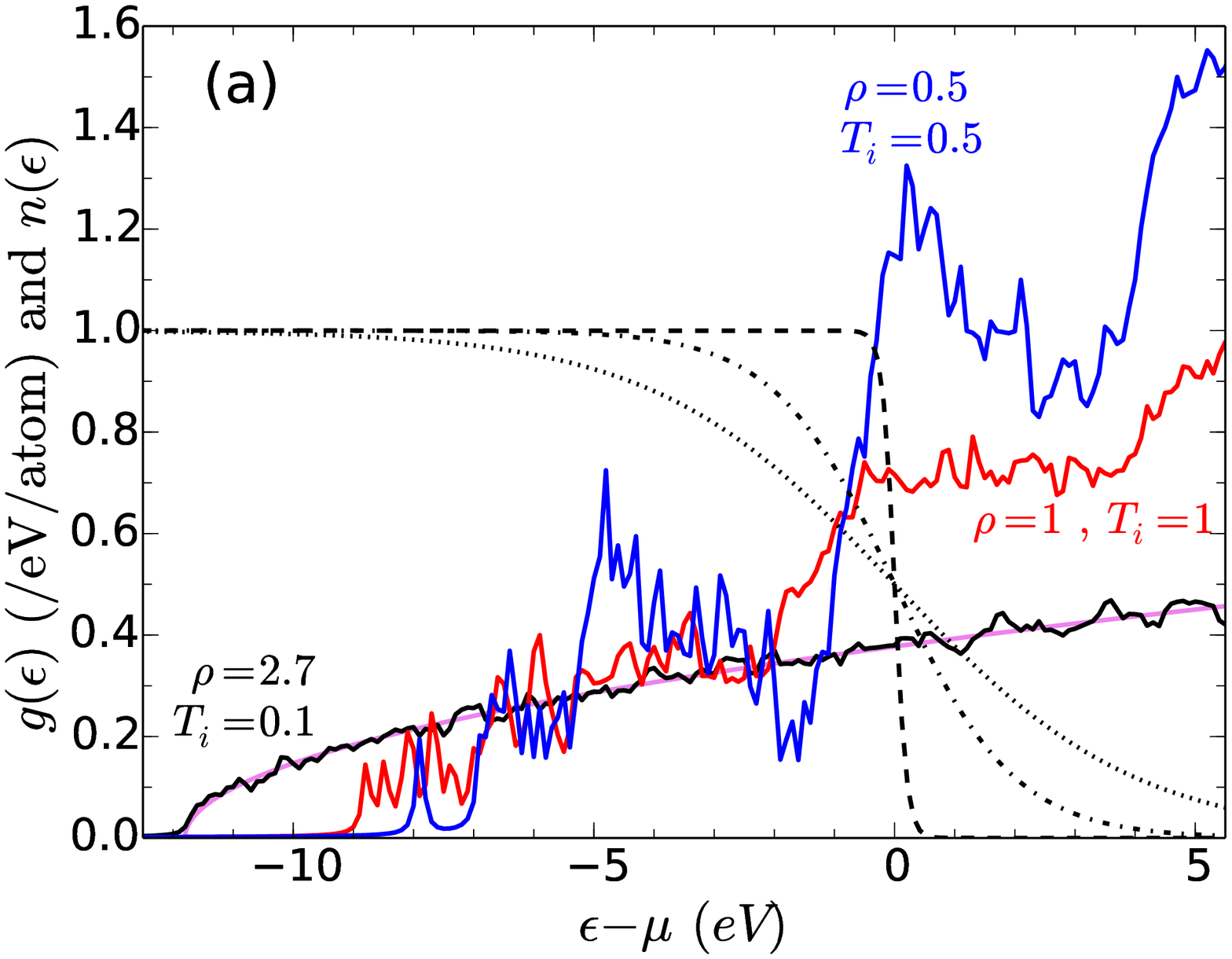}
\includegraphics[scale=0.36]{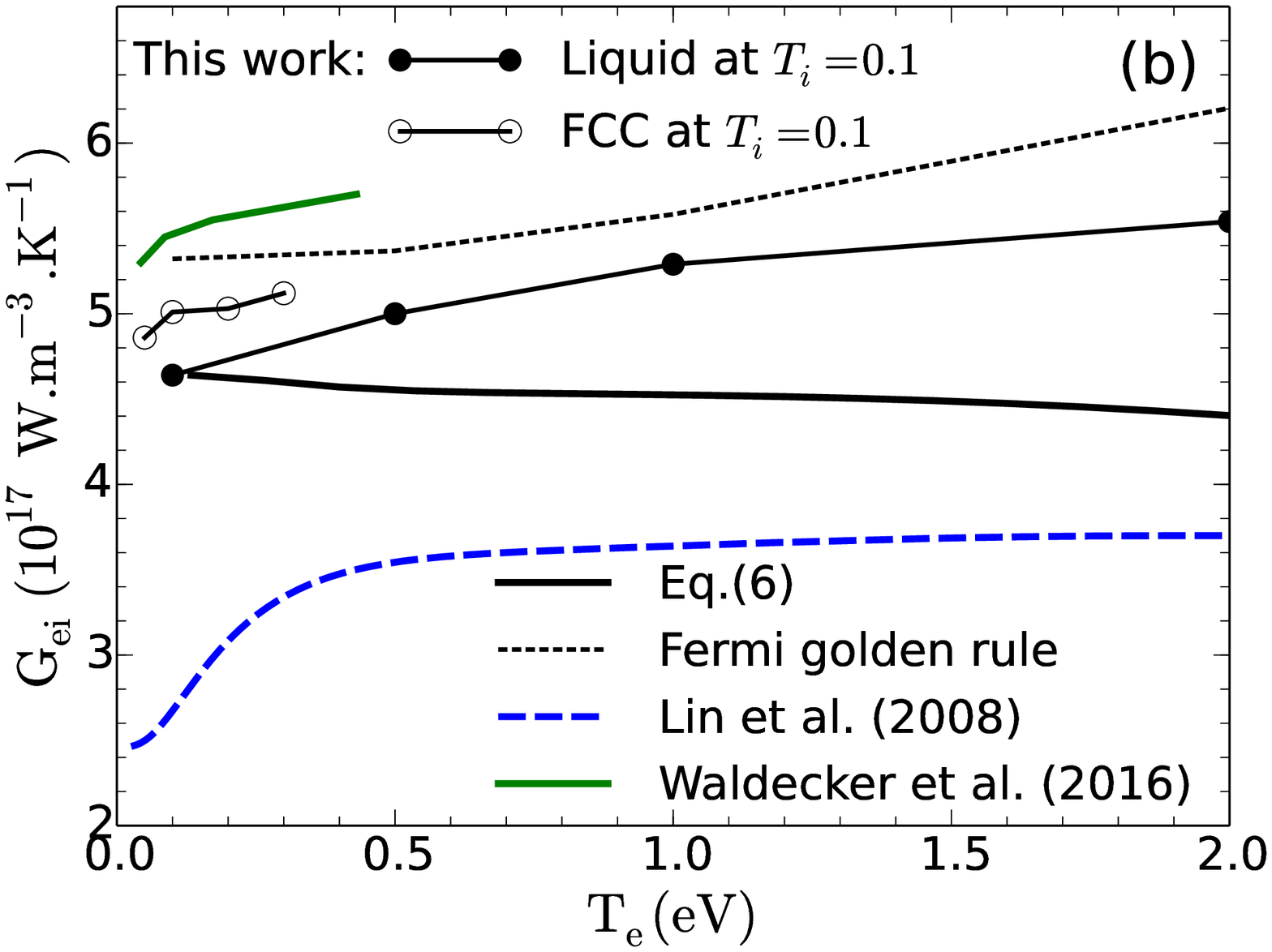}
\includegraphics[scale=0.36]{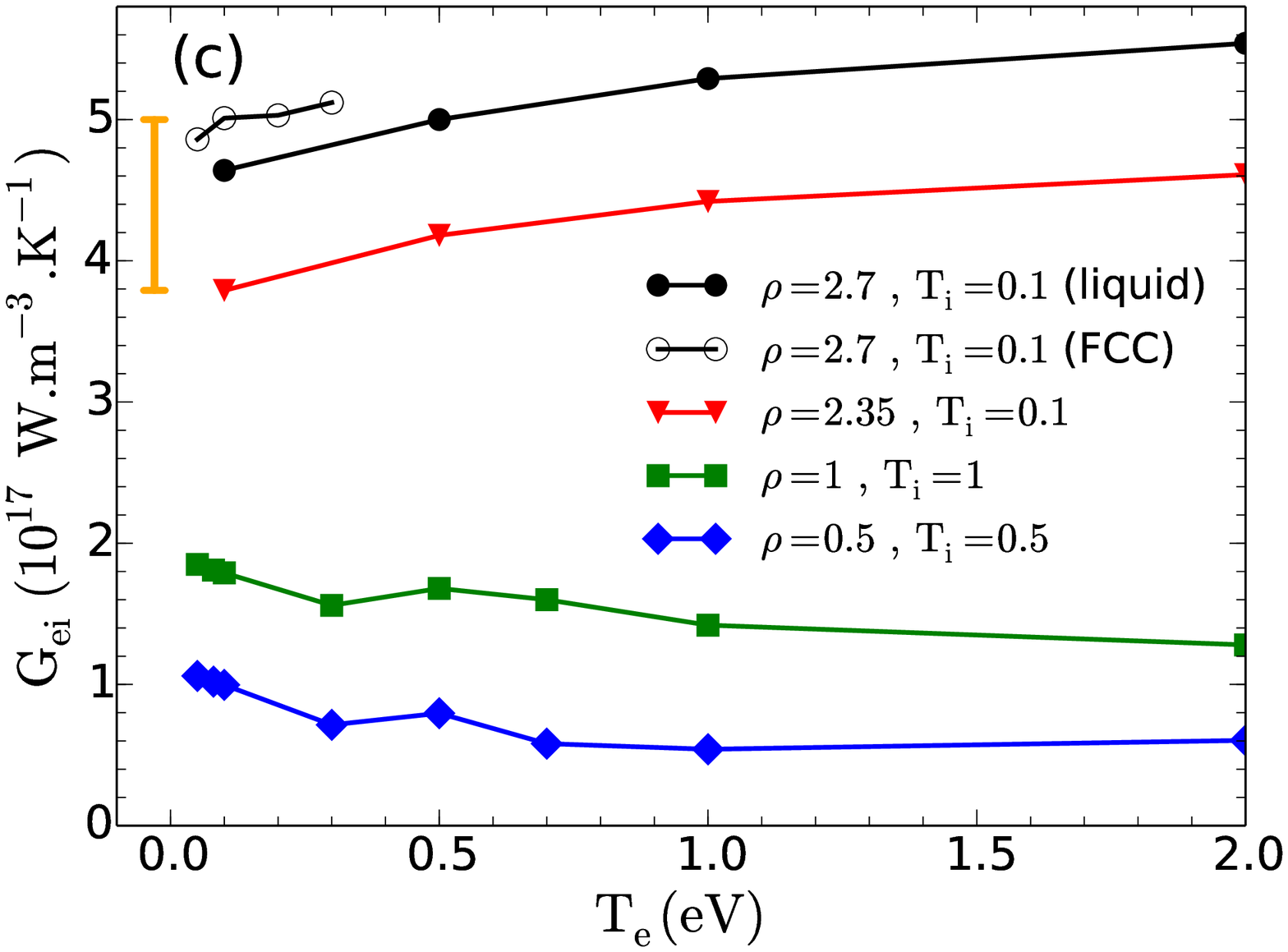}
\caption{(color online) (a) DOS of Al with $T_i=T_e$ and Fermi-Dirac distribution (dashed lines) for three electronic temperatures at $T_e=0.1,1,2$ eV. The violet line is the DOS of the free-electron gas at $2.7$ $\rm g.cm^{-3}$. Energy is measured with respect to the chemical potential $\mu(\rho,T_e)$.
(b) $G_{ei}(T_e,T_i)$ vs $T_e$ for solid density Al at $T_i=0.1$ eV compared with other model predictions (see all table~\ref{table_1}).
(c) $G_{ei}(T_e,T_i)$ vs $T_e$ for Al at various densities and ionic temperatures. The vertical bar indicates the magnitude of the variation of $G_{ei}$ at melting.
\label{figure_1}}
\end{figure}
\begin{figure*}[t]
\includegraphics[scale=0.29]{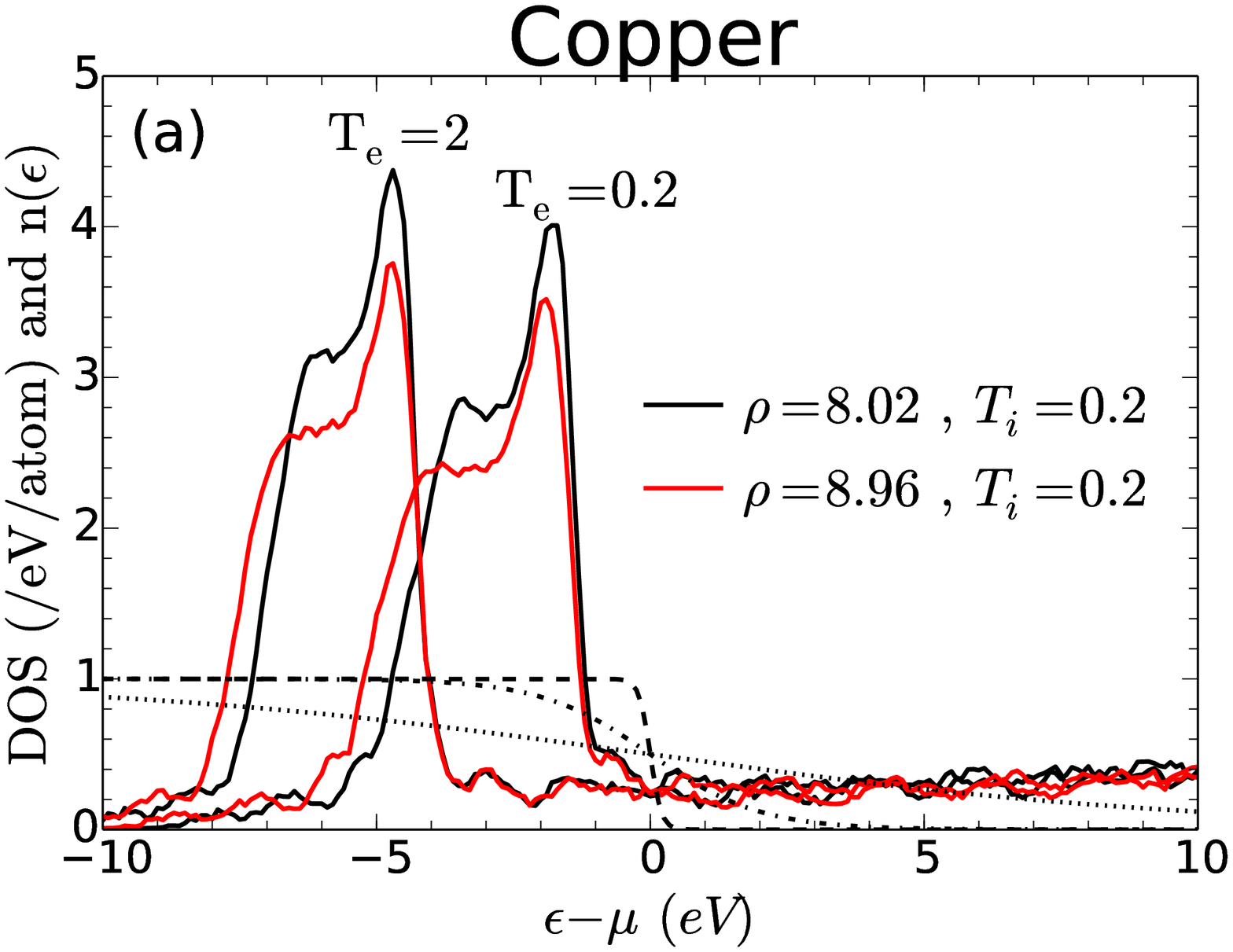}
\includegraphics[scale=0.29]{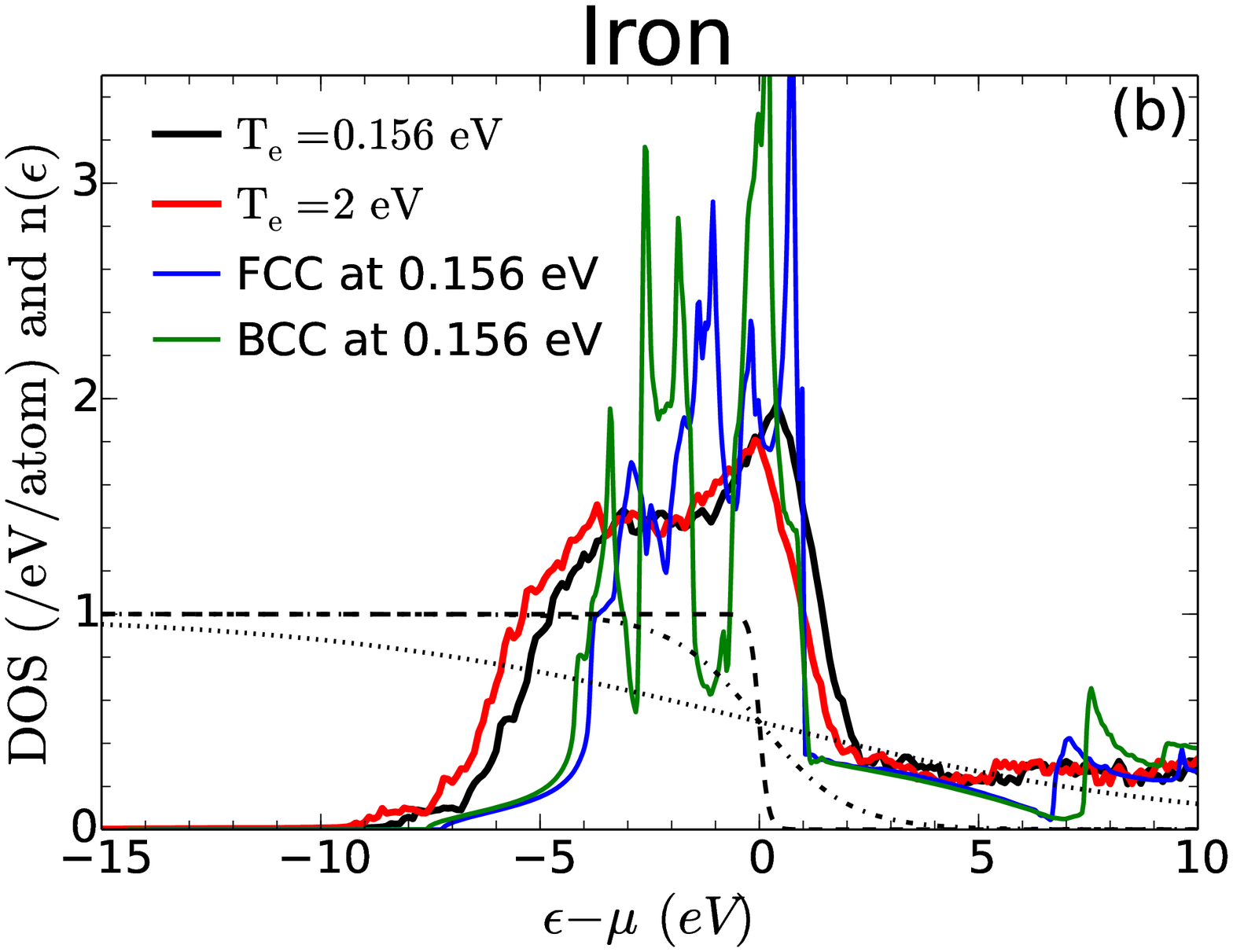}
\includegraphics[scale=0.29]{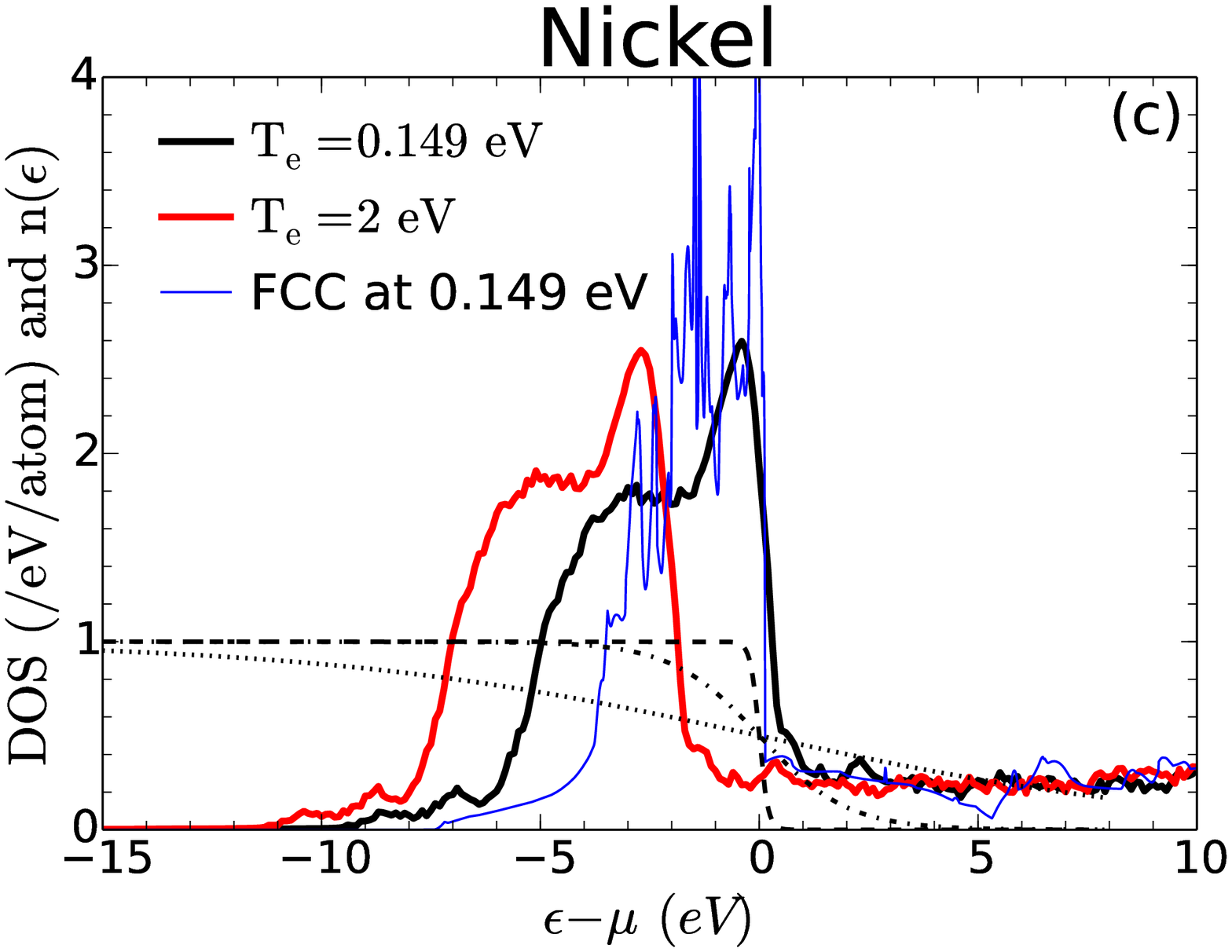}\\
\includegraphics[scale=0.29]{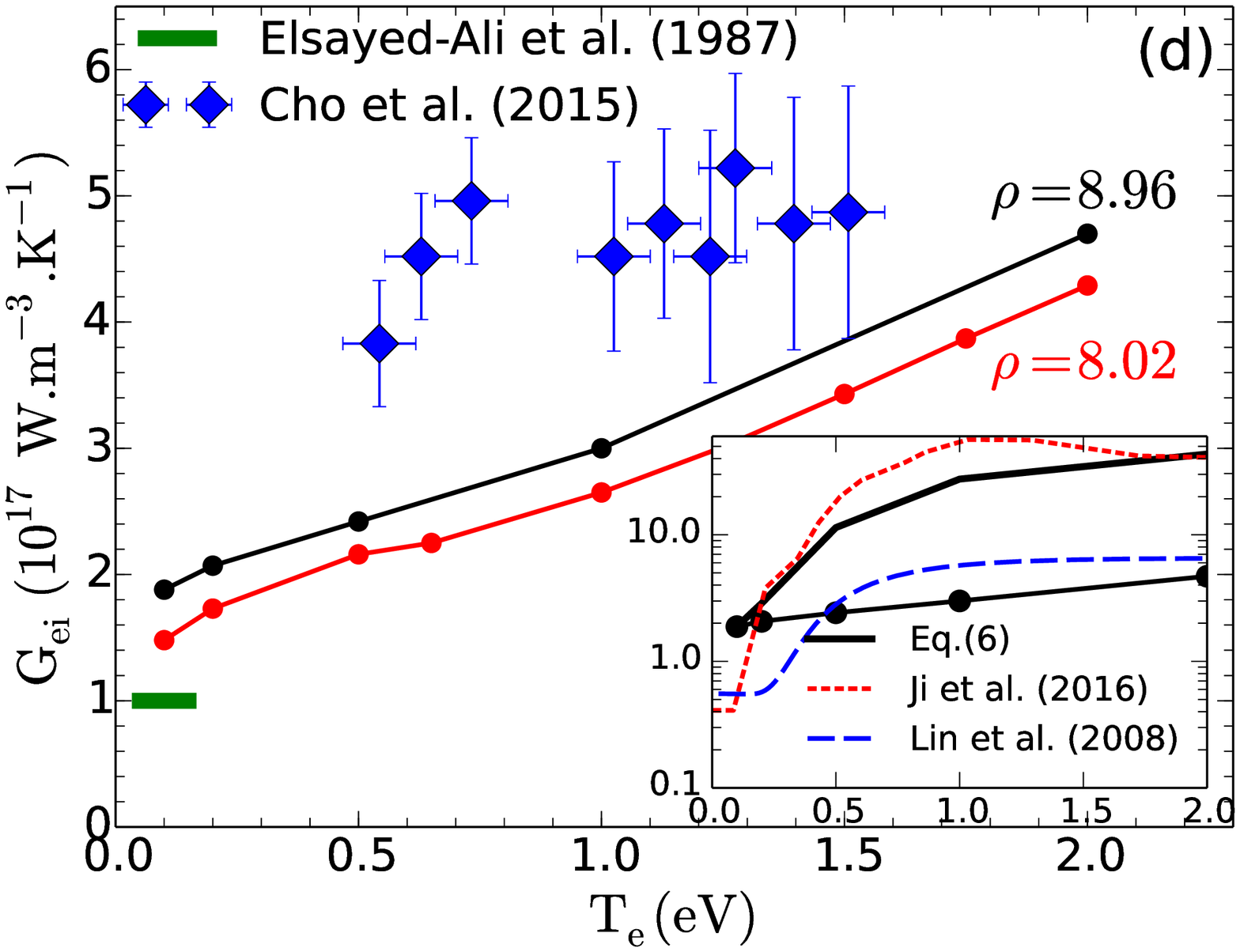}
\includegraphics[scale=0.29]{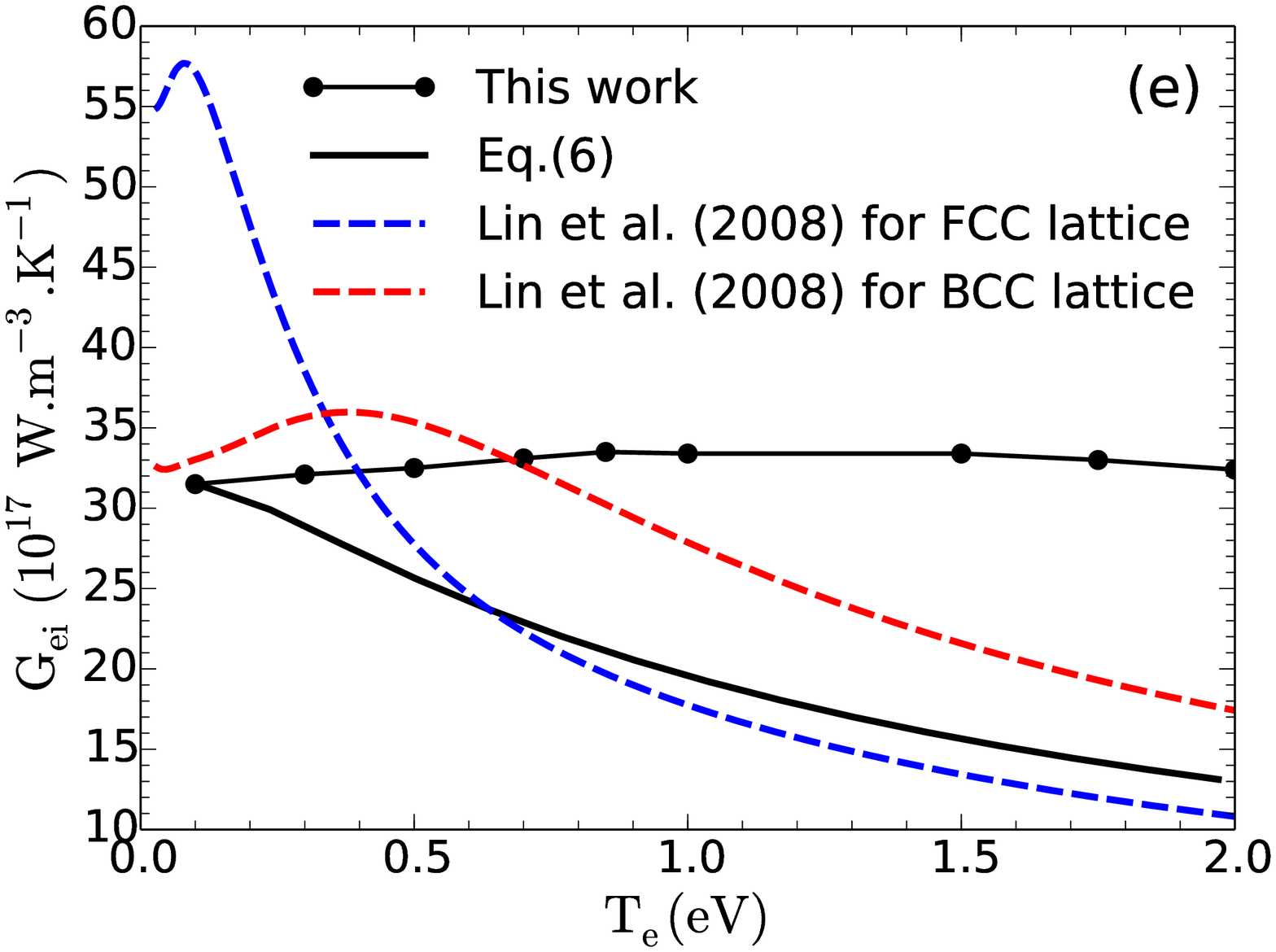}
\includegraphics[scale=0.29]{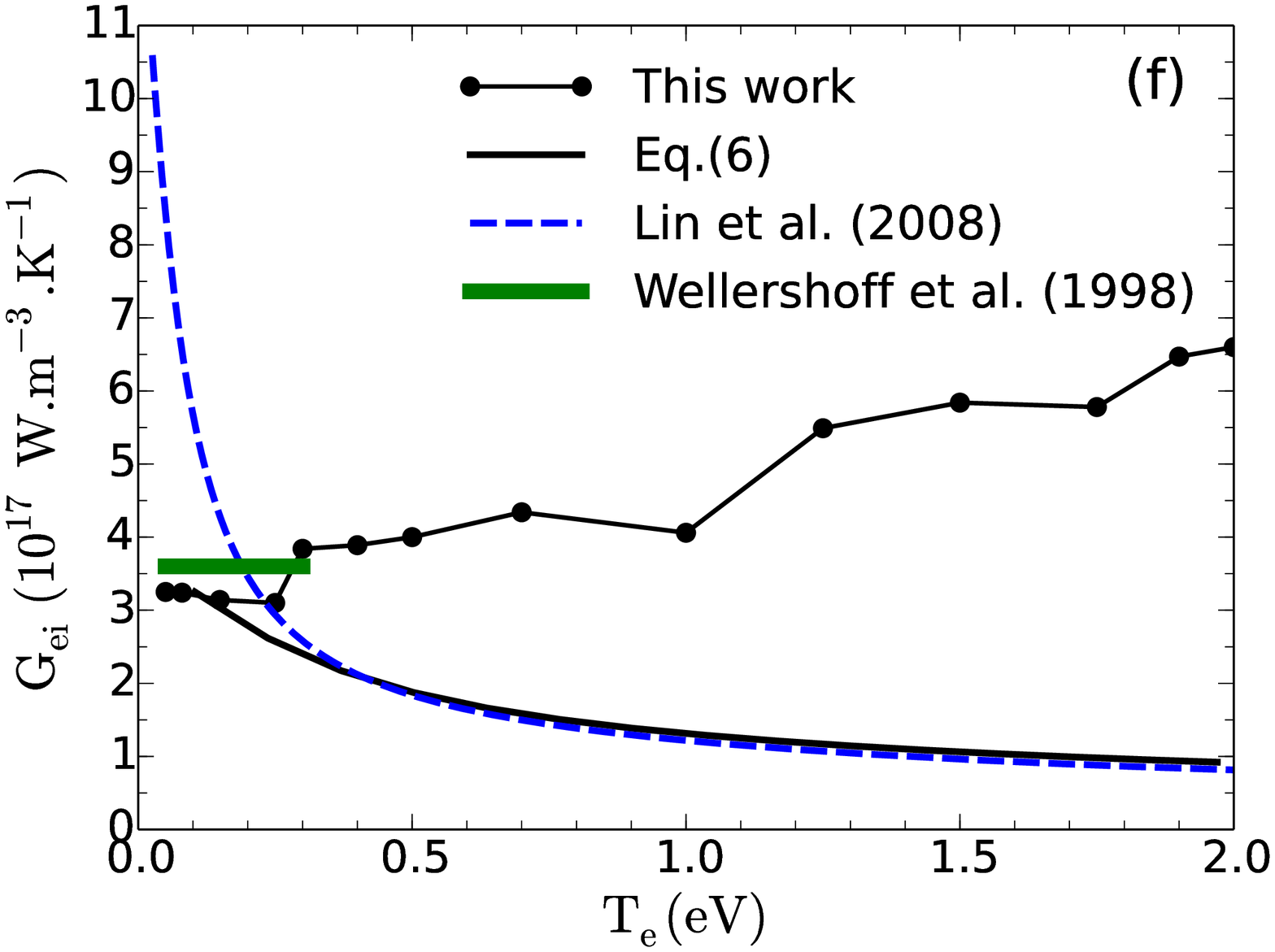}
\caption{(color online) Top panels: same as Fig.~\ref{figure_1}a for Cu, Fe and Ni at the conditions indicated in the legends.
Bottom panels: $G(T_e,T_i)$ vs $T_e$ for (d) solid and liquid density Cu at $T_i=0.2$ eV, (e) solid density Fe at $T_i=0.156$ eV, and (f) solid density Ni at $T_i=0.149$ eV.
In each case, the full lines with circles show the work's results, the full lines without symbols are obtained using Eq.(\ref{our_Lin_et_al}) with $G_0^{ei}$ set to reproduce the lowest $T_e$ value, the long dashed lines are the results based on Eq.(\ref{Lin_et_al}) discussed in \cite{Lin_et_al_2008}.
In panel (d), the diamonds show the experimental results of \cite{Cho_et_al_2015}, the bold green segment shows the measurement of \cite{Elsayed_et_al_1987} for solid Cu. In the inset, the dashed lines show the model predictions based on Eq.(\ref{Lin_et_al}) presented in \cite{Lin_et_al_2008, JiZhang2016}.
In panel (f), the bold green segment shows the measurement of \cite{Wellershoff1998}.
\label{figure_2}}
\end{figure*}
By following techniques similar to those used for the ab-initio calculation of electronic conductivities\cite{Holst2011}, we use the ionic and electronic structures calculated with standard quantum molecular dynamics simulations to evaluate the Kubo relations (\ref{gamma_IxJy_exact}) needed in Eq.(\ref{Gei}).
Briefly, for each ionic configuration ${\bf R}$, the electronic structure is obtained from the solution of the Kohn-Sham equations
$\big{(}\frac{\hat{{\bf p}}^2}{2m_e}+V_{KS}[\rho_e,{\bf R}]\big{)}|\alpha\rangle=\epsilon_\alpha |\alpha\rangle$, where
$\epsilon_\alpha$ and $|\alpha\rangle$ are the single-particle Kohn-Sham energies and states, $\rho_e({\bf r})=\sum_\alpha{n_\alpha|\langle{\bf r}|\alpha\rangle|^2}$ is the electron density, and $n_\alpha=n(\epsilon_\alpha)$ with $n(\epsilon)=\left(1+e^{-(\mu-\epsilon)/k_BT_e}\right)^{-1}$ represents the Fermi-Dirac occupation number of the state $\alpha$.
In terms of the Kohn-Sham quantities, it can be shown that the coupling coefficients (\ref{gamma_IxJy_exact})
\be
\gamma_{Ix,Jy}&=&
-\frac{\pi}{m_i}{\sum_{\alpha,\beta}}^\prime{\frac{n_\alpha-n_\beta}{\epsilon_{\alpha\beta}}f_{Ix}^{\alpha\beta}f_{Jy}^{\beta\alpha}\delta\left(\epsilon_{\alpha\beta}/\hbar\right)}\,, \label{gamma_IxJy_KS}
\ee
where the matrix elements $f_{Ix}^{\alpha\beta}=\left\langle \alpha \big| \hat{f}_{Ix}^{(sc)}\big|\beta\right\rangle$ and $\hat{f}_{Ix}^{(sc)}$ is the effective force along the $x$-direction between ion $I$ and a Kohn-Sham electron screened by the other electrons.

Before showing results, we relate our approach to a model that has served as a reference in recent works,
\be
G_{\rm e-ph}\approx G_0^{\rm e-ph}\!\!\int\limits_{-\infty}^{\infty}{\left[\frac{g(\epsilon)}{g(\epsilon_F)}\right]^2\!\!\left(\!-\frac{\partial\,n(\epsilon)}{\partial \epsilon}\right) \!d\epsilon}\,, \label{Lin_et_al}
\ee
which is a simplification in the high temperature limit \cite{Wang1994,Lin_et_al_2008} of the general electron-phonon coupling fomula \cite{Allen1987}.
Here $g(\epsilon)$ is the electron density of states (DOS), which is computable with DFT, and $G_0^{\rm e-ph}=\pi\hbar k_B\lambda\langle\omega^2\rangle g(\epsilon_F)$, where $\epsilon_F=k_BT_F$ is the Fermi energy, $\langle\omega^2\rangle$ is the second moment of the phonon spectrum, and $\lambda$ is the electron-phonon mass enhancement factor.
In previous works, the prefactor $G_0^{\rm e-ph} $ was either set to match an experimental measurement at low electronic temperature \cite{Lin_et_al_2008}, or was calculated ab-initio \cite{Waldecker2016, JiZhang2016}.
Although derived for crystalline solids, the model (\ref{Lin_et_al}) was used in recent works on warm dense matter systems \cite{Leguay2013,Cho_et_al_2015,Dorchies2016,Jourdain_et_al_2018,Ogitsu2018}.
Remarkably, an expression similar to Eq.(\ref{Lin_et_al}) also results from Eq.(\ref{gamma_IxJy_KS}) if one assumes that the matrix elements $f_{Ix}^{\alpha\beta}$ depend weakly on the energies, $f_{Ix}^{\alpha\beta}\approx f_{Ix}$, which yields
\be
G_{ei}\approx G_0^{ei} \int\limits_{-\infty}^{\infty}{\left[\frac{g(\epsilon)}{g(\epsilon_F)}\right]^2\!\!\left(\!-\frac{\partial\,n(\epsilon)}{\partial \epsilon}\right) d\epsilon}\,, \label{our_Lin_et_al}
\ee
where $G_0^{ei}=|f_{Ix}|^2g(\epsilon_F)^2$.
The formulas (\ref{Lin_et_al}) and (\ref{our_Lin_et_al}) highlights the interplay between the DOS and the distribution of electronic states, which, as shown by Lin et al. \cite{Lin_et_al_2008}, results in a strong dependence on the chemical composition and often on sharp variations with $T_e$.
Below we compare our results to predictions based on (\ref{Lin_et_al}) reported by others and on Eq.(\ref{our_Lin_et_al}) with $G_0^{ei}$ set to reproduce the value of $G_{ei}$ at the lowest $T_e$ considered.
We find that the simplified models (\ref{Lin_et_al}) and (\ref{our_Lin_et_al}) tend to overestimate the dependence on $T_e$ or predicts variations at odds with the full calculation.

Figures~\ref{figure_1} and \ref{figure_2} (bottom panels) show results for $G_{ei}(T_e,T_i)$ for five representative materials and physical conditions, together with the predictions of previous models and with experimental data.
Below we highlight some of the key findings.
For each element, the upper panels show the electron density of states $g(\epsilon)$ and the Fermi-Dirac distribution function $n(\epsilon)$ at representative conditions.
Our results were obtained with the open-source Quantum Espresso program \cite{QuantumEspressoCode}; the simulation details are given in the Supplemental Material \cite{SM}. 
In all cases, the material is prepared in the disordered, liquid-like state, except for Aluminum for which we also show calculations in a finite-temperature FCC configuration.
In the figures, temperatures $T_{i,e}$ are in eV and the material densities $\rho$ are in $\rm g.cm^{-3}$.

{\it Aluminum.}
Figure~\ref{figure_1}b shows $G_{ei}(T_i,T_e)$ versus $T_e$ at solid density $\rho\!=\! 2.7$ $\rm g.cm^{-3}$ and $T_i=0.1$ eV (slightly above the melting temperature $0.08$ eV), together with other model predictions, including the Fermi golden rule evaluated using the same pseudopotential $v_{ie}$ of the ab-initio calculations, and predicitons based on Eq.(\ref{our_Lin_et_al}) and the results of \cite{Lin_et_al_2008} and \cite{Waldecker2016} based on Eq.(\ref{Lin_et_al}) (see table~\ref{table_1} for other predictions).
$G_{ei}$ steadily increases between $4.6$ to $5.6$ $10^{17}\,\rm W/Km^{3}$ in the range $0.1\le T_e\le 2$ eV, as a result of the growing number of excited electrons that participate to the electron-ion scattering processes.
Our results are in best agreement with the Fermi-golden rule, which is expected given the free electron-like character of Al at solid density (see full black and violet lines in Fig.~\ref{figure_1}).
They differ from the prediction based on Eq.(\ref{our_Lin_et_al}), which is similar to the result one obtains with the DOS of the free-electron gas at solid density  (see Fig. 1d in \cite{Lin_et_al_2008}).
Figure~\ref{figure_1}c shows $G_{ei}$ at other mass densities $\rho$ and ionic temperatures $T_i$.
As $\rho$ decreases, the DOS shown in Fig.~\ref{figure_1}a progressively loses its free electron-like character.
We find that the $G_{ei}$ decreases with $\rho$ at constant $T_e$, which is essentially an effect of the variation of the decreasing electron density (see $n_e$ prefactor in Eq.(\ref{Gei})), and its variation with $T_e$ changes from an overall increasing to a decreasing functions of $T_e$.
The figures also show calculations obtained for FCC lattices at solid density (open circles in Fig~\ref{figure_1}b and c).
Our results are in good agreement with the result of Waldecker et al. \cite{Waldecker2016} based on Eq.(\ref{Lin_et_al}) with a DFT calculation of $G_0^{\rm e-ph}$.
At melting, the density is known to decrease from $\sim 2.7$ to $\sim 2.35$ $\rm g.cm^{-3}$ \cite{Leitner2017} and $G_{ie}$ decreases by about $25\%$, as indicated by the orange vertical bar in Fig.~\ref{figure_1}c.
This should be contrasted with the large change in the electrical resisitivity at melting, which increases by a factor $\sim 2.1$ \cite{Leitner2017}, in other words disorder has a higher effect on momentum relaxation than on energy relaxation.

{\it Copper.}
Warm dense copper has been the focus of several recent studies \cite{Cho_et_al_2015,Jourdain_et_al_2018,Lin_et_al_2008,JiZhang2016}.
Figure~\ref{figure_2}d shows results at solid and melt densities, $8.96$ and $8.02$ $\rm g.cm^{-3}$, and $T_i=0.2$ eV (melting temperature is $0.117$ eV), together with the measurements of \cite{Elsayed_et_al_1987} and \cite{Cho_et_al_2015}; the inset compares our result at $8.96$ $\rm g.cm^{-3}$ with Eq.(\ref{our_Lin_et_al}) and with the results of \cite{Lin_et_al_2008} and \cite{JiZhang2016} based on Eq(\ref{Lin_et_al}).
We find that $G_{ei}$ increases with $T_e$, with a faster variation above $0.5$ eV when the $d$ electrons, which are responsible for the prominent regions of high DOS in Fig.~\ref{figure_2}a, can be excited and participate the electron-ion energy exchanges.
However, the variation is not as sharp and intense as that predicted using Eq.(\ref{Lin_et_al}) of \cite{Lin_et_al_2008} and \cite{JiZhang2016}. 
Unlike Ref.~\cite{Lin_et_al_2008}, we don't find a sharp increase of $G_{ei}$ at small $T_e$, which was ascribed to the thermal excitations of d-electrons.
At solid density, we find $G_{ei}\simeq 2$ $10^{17}\,\rm W/Km^{3}$, in fair agreement with the old measurement $10^{17}\,\rm W/Km^{3}$ of Elsayed-Ali et al. \cite{Elsayed_et_al_1987} for solid Cu.
Our data lie slightly below the recent measurements reported in \cite{Cho_et_al_2015}.

{\it Iron.}
Figure~\ref{figure_2}e shows the variation of $G_{ei}$ with $T_e\le 2$ eV for solid density Fe $\rho=7.87$ $\rm g.cm^{-3}$ at melting temperature $T_i=0.156$ eV.
We find that $G_{ei}$ does not vary significantly over the temperature range considered, unlike the predictions based on Eqs.(\ref{Lin_et_al}) \cite{Lin_et_al_2008} and (\ref{our_Lin_et_al}).

{\it Nickel.}
Figure~\ref{figure_2}f shows the variation of $G_{ei}$ with $T_e\le 2$ eV for solid density Ni $\rho=8.91$ $\rm g.cm^{-3}$ at melting temperature $T_i=0.149$ eV.
We find that $G_{ei}$ increases from $3.1$ to $5.6$ $10^{17}\,\rm W/Km^{3}$ over the temperature range, in contrast with the results based on Eq.(\ref{Lin_et_al}) \cite{Lin_et_al_2008} and on Eq.(\ref{our_Lin_et_al}).
Our result at lower $T_e$ are in good agreement with the measurement reported by Wellershoff et al. \cite{Wellershoff1998}.

In summary, we have presented much-needed first-principles calculations of the electron-ion coupling factors of materials at the confluence of solids and plasmas based on a general expression in terms of the friction coefficients felt by ions due to the non-adiabatic electron-ion interactions.
The approach serves as a useful comparison with the experimental measurements, permits an extension into conditions not covered by experiments, and provides insight into the underlying physics.
We hope that this work will help assist and motivate future experiments and, ultimately, will help advance our understanding of the warm dense matter regime.

\begin{acknowledgments}
This work was supported by the US Department of Energy through the Los Alamos National Laboratory through the LDRD Grant No. 20170490ER and the Center of Non-Linear Studies (CNLS).
Los Alamos National Laboratory is operated by Triad National Security, LLC, for the National Nuclear Security Administration of U.S. Department of Energy (Contract No. 89233218CNA000001).
\end{acknowledgments}

\end{document}

\title{Supplemental Material for 'First-Principles Determination of Electron-Ion Couplings in the Warm Dense Matter Regime'}
\author{Jacopo Simoni}\email[Contact email address: ]{jsimoni@lanl.gov}
\affiliation{Theoretical Division, Los Alamos National Laboratory, Los Alamos, NM 87545} 
\author{J\'erome Daligault} 
\affiliation{Theoretical Division, Los Alamos National Laboratory, Los Alamos, NM 87545}

\pacs{75.75.+a, 73.63.Rt, 75.60.Jk, 72.70.+m}

\maketitle

\section{The friction tensor for a periodic system}

In order to compute the electron-ion coupling of materials under different conditions of temperature and density we combine a finite temperature Density Functional Theory\cite{DFT1,DFT2} (FT-DFT) approach that allows us to estimate the ground state of the electronic system at any given atomic configuration together with classical Molecular Dynamics (MD) simulations to temporally evolve the ionic positions. Since in order to compute the temperature relaxation we need to evaluate a thermal average over the classical ionic degrees of freedom, the system should be let evolved for a time sufficient to collect enough configurations. The ground state of the electronic system is obtained by solving the following set of Kohn-Sham (KS) equations\cite{KS}
\begin{equation}\label{Eq:KS}
  \hat{H}_\mathrm{KS}\ket{\alpha,\mathbf{k}} = \epsilon_\alpha(\mathbf{k})\ket{\alpha,\mathbf{k}},
\end{equation}
where $\hat{H}_\mathrm{KS}$ is the KS Hamiltonian, $\ket{\alpha,\mathbf{k}}$ represents the effective single particle Kohn-Sham state with energy eigenvalue $\epsilon_\alpha(\mathbf{k})$. The KS Hamiltonian written on a spatial grid acquires the following form
\begin{eqnarray}
  H_\mathrm{KS}(\mathbf{r}) & = & -\frac{\hbar^2\boldsymbol{\nabla}_\mathbf{r}^2}{2m_\mathrm{e}} + v_\mathrm{KS}(\mathbf{r}),\\
  v_\mathrm{KS}(\mathbf{r}) & = & v_\mathrm{H}(\mathbf{r}) + v_\mathrm{xc}(\mathbf{r}) + v_\mathrm{ext}(\mathbf{r}),
\end{eqnarray}
where the KS potential $v_\mathrm{KS}(\mathbf{r})$ is given by the sum of the Hartree component, $v_\mathrm{H}(\mathbf{r})$, the exchange-correlation component, $v_\mathrm{xc}(\mathbf{r})$, and the external ionic potential $v_\mathrm{ext}(\mathbf{r})$.

The electron-ion coupling in the warm dense matter regime can be obtained from the calculation of the friction coefficients $\gamma_\mathrm{I,x}(\{\mathbf{R}_\mathrm{I}\},T_\mathrm{e})$. These coefficients are written in Eq.~(3) of the main paper, and it can be shown (the details will be presented in~(\onlinecite{longpaper})) that they are equivalently expressed in terms of the following Kubo-Greenwood expression written here in the case of periodic systems
\begin{equation}
  \gamma_\mathrm{I,x}(\{\mathbf{R}\},T_\mathrm{e}) = -\frac{1}{m_\mathrm{i}}\int_0^\infty\,dt\sum_{\alpha\neq\beta}\sum_{\mathbf{k}\in\mathrm{IBZ}}W_\mathbf{k}\frac{n_\alpha(\mathbf{k})-n_\beta(\mathbf{k})}{\epsilon_\alpha(\mathbf{k})-\epsilon_\beta(\mathbf{k})}f_{\alpha\beta}^\mathrm{Ix}(\mathbf{k})f_{\beta\alpha}^\mathrm{Ix}(\mathbf{k})\cos\bigg(\frac{\epsilon_\alpha(\mathbf{k})-\epsilon_\beta(\mathbf{k})}{\hbar}t\bigg).
\end{equation}
In the previous quantity the first summation is computed over all the possible transitions between the computed KS bands $\alpha$ and $\beta$ with $\alpha\neq\beta$ and the second over all the {\bf k}-points belonging to the Irreducible Brillouin Zone (IBZ). $n_\alpha(\mathbf{k})=(1 + e^{-\beta_\mathrm{e}(\mu-\epsilon_\alpha(\mathbf{k}))})^{-1}$ is the Fermi-Dirac occupation for the KS state $\ket{\alpha,\mathbf{k}}$ and $W_\mathbf{k}$ defines the {\bf k}-point integration weights.

The force matrix elements $f_{\alpha\beta}^\mathrm{Ix}(\mathbf{k})$ are obtained from the following integral in real space
\begin{equation}\label{Eq:1}
  f_{\alpha\beta}^\mathrm{Ix}(\mathbf{k}) = \int_\Omega\dd[3]{r}u_{\alpha\mathbf{k}}(\mathbf{r})^* \boldsymbol{\nabla}_{r_x}v_\mathrm{Ie}^\mathrm{scr}(\mathbf{r}-\mathbf{R}_\mathrm{I}) u_{\beta\mathbf{k}}(\mathbf{r}),
\end{equation}
where $u_{\alpha\mathbf{k}}(\mathbf{r})$ is the Bloch component of the full Kohn-Sham wave function $\Psi_{\alpha\mathbf{k}}(\mathbf{r})=\frac{1}{\sqrt{N_\mathrm{c}}}u_{\alpha\mathbf{k}}(\mathbf{r})e^{i\mathbf{k}\cdot\mathbf{r}}$, $v_\mathrm{Ie}^\mathrm{scr}(\mathbf{r}-\mathbf{R}_\mathrm{I})$ is the effective screened electron-ion pseudo-potential centered on atom $I$, $\Omega$ is the volume of the unit cell and $N_\mathrm{c}$ is a normalization factor for the system's wave functions $\Psi_{\alpha\mathbf{k}}(\mathbf{r})$ equivalent to the total number of {\bf k} points used in the calculation.

Some more details are required in order to compute the screened electron-ion potential gradient, $\boldsymbol{\nabla}_{r_x}v_\mathrm{Ie}^\mathrm{scr}(\mathbf{r}-\mathbf{R}_\mathrm{I})$, appearing in the previous expression. In the case when a local pseudo-potential, $v_\mathrm{Ie}$, is used to describe the interaction between the ions and the electrons, it has to be screened through the KS dielectric function, $\varepsilon_\mathrm{KS}$, according to the following equation in order to account for the presence of the other electrons
\begin{equation}
  \boldsymbol{\nabla}_{r_x}v_\mathrm{Ie}^\mathrm{scr}(\mathbf{r}-\mathbf{R}_\mathrm{I}) = \int_V\dd[3]{r'} \boldsymbol{\nabla}_{r'_x}v_\mathrm{Ie}(\mathbf{r'}-\mathbf{R}_\mathrm{I}) \varepsilon_\mathrm{KS}(\mathbf{r'},\mathbf{r},\omega=0),
\end{equation}
however, in the case of projected augmented wave potentials\cite{PAW} the situation is different and Eq.~(\ref{Eq:1}) needs to be modified in order to account for the corrections on the valence pseudo wave functions close to the atomic nuclei
\begin{equation}\label{Eq:2}
  f_{\alpha\beta}^{\mathrm{I}x}(\mathbf{k}) = \mel*{\tilde{\Psi}_{\alpha\mathbf{k}}}{\boldsymbol{\nabla}_{r_x}v_\mathrm{AE}^\mathrm{scr}(\mathbf{r}-\mathbf{R}_\mathrm{I})}{\tilde{\Psi}_{\beta\mathbf{k}}} + \sum_{a=1}^N\sum_{n,m}\ip*{\tilde{\Psi}_{\alpha\mathbf{k}}}{\tilde{p}_{n,\mathbf{R}_a}}\Delta f_{nm}^a\ip*{\tilde{p}_{m,\mathbf{R}_a}}{\tilde{\Psi}_{\beta\mathbf{k}}},
\end{equation}
here $\tilde{\Psi}_{\alpha\mathbf{k}}$ is the smooth pseudo wave function for the state $\ket{\alpha,\mathbf{k}}$, and the potential used in order to compute the forces is now the screened all-electron potential of the periodic system centered on atom $\mathrm{I}$, $v_\mathrm{AE}^\mathrm{scr}(\mathbf{r}-\mathbf{R}_\mathrm{I})=-\sum_{l,\boldsymbol{\tau}}\frac{Z e^2}{4\pi\epsilon_0|\mathbf{r}-\mathbf{R}_\mathrm{I}+l\boldsymbol{\tau}|}$, where $\boldsymbol{\tau}$ are the unit cell's lattice vectors, now the screening should not account only for the valence electrons but also for the core electrons that are frozen around the nuclei, therefore the dielectric function needs to be modified accordingly
\begin{equation}
  \boldsymbol{\nabla}_{r_x}v_\mathrm{AE}^\mathrm{scr}(\mathbf{r}-\mathbf{R}_\mathrm{I}) = \int_V\dd[3]{r'}\boldsymbol{\nabla}_{r'_x}v_\mathrm{AE}(\mathbf{r'}-\mathbf{R}_\mathrm{I})\varepsilon_\mathrm{KS}^\mathrm{AE}(\mathbf{r'},\mathbf{r},\omega=0).
\end{equation}
Back to Eq.~(\ref{Eq:2}), $\ket{\tilde{p}_{n,\mathbf{R}_a}}$ represents the projector over the valence atomic state $\ket{\phi_{n,\mathbf{R}_a}}$ centered on the atomic sphere $a$, while
\begin{equation}
  \Delta f_{nm}^a = \mel{\phi_{n,\mathbf{R}_a}}{\boldsymbol{\nabla}_{r_x}v_\mathrm{AE}^\mathrm{scr}(\mathbf{r}-\mathbf{R}_\mathrm{I})}{\phi_{m,\mathbf{R}_a}}-\mel*{\tilde{\phi}_{n,\mathbf{R}_a}}{\boldsymbol{\nabla}_{r_x}v_\mathrm{AE}^\mathrm{scr}(\mathbf{r}-\mathbf{R}_\mathrm{I})}{\tilde{\phi}_{m,\mathbf{R}_a}},
\end{equation}
defines the core correction matrix elements, where $\ket{\tilde{\phi}_{n,\mathbf{R}_a}}$ is the pseudo valence atomic wave function centered on atom $a$.  

\section{Details of the calculations}
The temperature relaxation rate is given by the following thermal average
\begin{equation}
  g(T_\mathrm{e}) = \frac{3k_\mathrm{B}n_\mathrm{ion}}{N_\mathrm{conf}}\sum_{\{\mathbf{R}\}}\Bigg[\frac{1}{3N_\mathrm{at}}\sum_{\mathrm{I}=1}^{N_\mathrm{at}}\sum_{x=1}^3\gamma_{\mathrm{I},x}(\{\mathbf{R}\},T_\mathrm{e})\Bigg],
\end{equation}
with $n_\mathrm{ion}$ ionic number density, $k_\mathrm{B}$ Boltzmann constant and $N_\mathrm{conf}$ total number of configurations considered in the average. For each set of atomic positions $\mathbf{R}$ we compute the electronic ground state by applying FT-DFT at the level implemented in the QUANTUM ESPRESSO $5.1$ program suite\cite{QE1,QE2}, the set of KS equations (\ref{Eq:KS}) are solved until convergence in the ground state electron density is reached. Then we can compute the forces acting on each ion and update the atomic positions at the successive MD step.

This procedure is executed repeatedly in a periodic simulation box of cubic shape, the time step used in the MD simulation is approximately \SI{0.5}{fs}, the system is then evolved for $3000$-$5000$ MD iterations in order to reach thermalization and then for other $5000$-$7000$ iterations to select different and well separated ionic configurations, the number of MD steps used may change with the number of atoms, temperature and density of the system. The ionic temperature is controlled with a Andersen thermostat\cite{And}, while The FT-DFT calculation is performed at a single $\Gamma$ point, the energy cut-off was set between \SI{200}{eV} and \SI{270}{eV} depending on the calculation. The number of bands employed depends strongly on the electronic temperature of the system, the higher the electronic temperature is, the higher the number of bands required in the calculation is in order to converge. The number of atoms used within the simulation box is also a very important parameter, in particular in the case of Aluminum's systems we always use $64$ atoms in the periodic box except at $\SI{0.5}{gr./cm^3}$ where we use $48$ atoms and in the case of the electron-phonon calculation with the FCC structure where we use $256$ atoms instead. For all the other chemical elements the simulations are performed with a periodic box of $64$ atoms.

For Aluminum and Iron systems we used the Perdew-Zunger\cite{LDA} (PZ) Local Density Approximation (LDA) to compute the exchange correlation potential, $v_\mathrm{xc}(\mathbf{r})$, while in the other cases we used a PBE type of exchange correlation potential\cite{PBE}.
The interaction potential between the electrons and the ions is handled by using either a local pseudo-potentials in the case of Aluminum at \SI{2.7}{gr/cm^3} and \SI{2.35}{gr/cm^3}, or by using the projector-augmented wave method in all the other cases as we have already mentioned in the previous section.

After that the system's thermalization has been reached we select a set of atomic configurations and then we perform FT-DFT calculations at different electronic temperatures for each of them. In practice, a single atomic configuration is sufficient in order to obtain a reasonably well converged temperature relaxation, this is due to the fact that all the atomic contributions to $\gamma_\mathrm{I,x}(\{\mathbf{R}\},T_\mathrm{e})$ are quite similar and contribute almost equally to the final sum, $g(T_\mathrm{e})$ becomes then a simple average of the diagonal elements of the friction tensor
\begin{equation}
  g(T_\mathrm{e}) = \frac{3k_\mathrm{B}n_\mathrm{ion}}{3N_\mathrm{at}}\sum_{\mathrm{I}=1}^\mathrm{N_{at}}\sum_{\mathrm{x}=1}^3\gamma_\mathrm{I,x}(\{\mathbf{R}\},T_\mathrm{e}).
\end{equation}
For this single ionic configuration the FT-DFT calculation is performed more accurately than during the MD iterations. Once we compute the electronic ground state and we obtain the full set of Kohn-Sham wave functions, $\{\Psi_{\alpha\mathbf{k}}(\mathbf{r})\}_{\alpha\mathbf{k}}$, the force matrix elements, $f_{\alpha\beta}^\mathrm{Ix}(\mathbf{k})$, are computed by using either Eq.~(\ref{Eq:1}) in the case of a local pseudo-potential or Eq.~(\ref{Eq:2}) with a projected augmented wave potential.
In order to achieve convergence a much higher number of bands was required depending on the number of atoms used in the simulation box and on the electronic temperature as we have mentioned earlier, however we always tried to keep the occupation number $n_\alpha(\mathbf{k})$ of the highest occupied energy states lower than a treshold value $\epsilon$, this treshold is set to a value $\epsilon\simeq 10^{-3}$ at high electronic temperatures, but it may be much lower in the case of low electronic temperatures when a smaller number of bands is required to achieve convergence. The energy cut-off was set in almost all the calculations to a value of \SI{2040}{eV} in order to have a sufficiently fine spatial grid for a better representation of the potential's gradient. We used $2\times 2\times 2$ number of {\bf k} points in almost all the calculations since it was giving already well converged values for the relaxation rate, but at higher densities a higher number of {\bf k} points may be required. In order to produce the density of states for Aluminum's systems at different mass densities shown in Fig.~(1a) of the main paper we used a total of $64$ {\bf k} points ($4\times 4\times 4$). 

%%%%%%%%%%%%%%%%%%

%%%%%%%%%%%%%%%%%%%%%%%%%%%%%%%%%%%%%